\input harvmac
\input epsf


\lref\rDor{N. Dorey, V.V. Khoze and M.P. Mattis, {\it On N=2 Supersymmetric
QCD with 4 Flavors,} Nucl. Phys. B492 (1997) 607 hep-th 9611016.}
\lref\rDL{I.L. Buchbinder and S.M. Kuzenko, {\it Comments on the background
field method in Harmonic Superspace: Non-holomorphic Corrections in N=4
SYM,} Mod. Phys. Lett. A13 (1998) 1623, hep-th 9804168\semi
E.I. Buchbinder, I.L. Buchbinder and S.M. Kuzenko, {\it Non-holomorphic
effective potential in N=4 U(n) SYM,} hep-th 9810239\semi 
D.A. Lowe and R. von Unge, {\it `Constraints on Higher Derivative
Operators in Maximally Supersymmetric Gauge Theory'}, hep-th 9811017.}
\lref\rHW{A. Hanany and E. Witten, {\it Type IIB Superstrings, BPS Monopoles
and three dimensional gauge dynamics,} Nucl. Phys. B492 (1997) 152-190,
hep-th 9611230\semi
E. Witten, {\it Solutions of four-dimensional field theories via
M theory,} Nucl. Phys. B500 (1997) 3-42, hep-th 9703166.}
\lref\rHol{A. Fayyazuddin and M. Spalinski, {\it The Seiberg Witten Differential
from M Theory,} Nucl. Phys. B508 (1997) 219-228, hep-th 9706087\semi 
M. Henningson and P. Yi, {\it Four Dimensional BPS Spectra via M theory,}
Phys. Rev. D57 (1998) 1291-1298, hep-th 9707251\semi
A. Mikhailov, {\it BPS States and Minimal Surfaces,} hep-th 9708068.}
\lref\rWest{N.D. Lambert and P.C. West, {\it N=2 Superfields 
and the M Fivebrane,} Phys. Lett. B424 (1998) 281-287, hep-th 9801104\semi
N.D. Lambert and P.C. West, {\it Gauge Fields and M Fivebrane Dynamics,}
Nucl. Phys. B524 (1998) 141-158, hep-th 9712040\semi
P.S. Howe, N.D. Lambert and P.C. West, {\it Classical M Fivebrane Dynamics and
Quantum N=2 Yang-Mills,} Phys. Lett. B418 (1998) 85-90\semi
J. de Boer, K. Hori, H. Ooguri and Y. Oz, {\it Kahler Potential and Higher 
Derivative Terms from M Theory Fivebranes,} Nucl. Phys. B518 (1998) 173-211,
hep-th 9711143.}
\lref\rM{O. Aharony, A. Fayyazuddin and J. Maldacena, {\it The Large N limit
of ${\cal N}=1,2$ Field Theories from Threebranes in F-Theory,} hep-th 9806159.}
\lref\rSW{N. Seiberg and E. Witten, {\it Electric-Magnetic Duality, Monopole
Condensation and Confinement in N=2 Supersymmetric Yang-Mills Theory,} Nucl.
Phys. B426 (1994) 19-52, Erratum-ibid B430 (1994) 485-486, hep-th 9407087;
{\it Monopoles, Duality and Chiral Symmetry Breaking in N=2 Supersymmetric
QCD,} Nucl. Phys. B431 (1994) 484-550, hep-th 9408099.}
\lref\rOL{S.V. Ketov, {\it On the Next-to-leading order correction to the
Effective Action in ${\cal N}=2$ gauge theories,} hep-th 9706079.}
\lref\rTL{I.L. Buchbinder, S.M. Kuzenko and B.A. Ovrut, {\it On the $D=4$
${\cal N}=2$ Non-Renormalization Theorem,} hep-th 9710142.}
\lref\rOI{A. Yung, {\it Higher Derivative Terms in the Effective Action
of ${\cal N}=2$ SUSY QCD from Instantons,} hep-th 9705181.}
\lref\rTI{D. Bellisai, F. Fucito, M. Matone and G. Travaglini, {\it 
Non-Holomorphic Terms in ${\cal N}=2$ SUSY Wilsonian Actions and RG
Equation,} hep-th 9706099.}
\lref\rYOz{J. de Boer, K. Hori, H. Ooguri and Y. Oz, 
{\it Kahler Potential and Higher 
Derivative Terms from M Theory Fivebranes,} Nucl. Phys. B518 (1998) 173-211,
hep-th 9711143.}
\lref\rEx{M. Henningson, {\it Extended Superspace, Higher Derivatives and 
$SL(2,Z)$ Duality,} hep-th 9507135\semi
B. de Wit, M.T. Grisaru and M. Rocek, {\it Nonholomorphic Corrections to the 
One Loop ${\cal N}=2$ Super Yang-Mills action,} hep-th 9601115.}
\lref\rEW{E. Witten, {\it Solutions of four-dimensional field theories via
M theory,} Nucl. Phys. B500 (1997) 3-42, hep-th 9703166.}
\lref\rJR{R. de Mello Koch and J.P. Rodrigues, {\it Solving Four Dimensional 
Field Theories with the Dirichlet Fivebrane,} hep-th 9811036.}
\lref\rTS{A.A. Tseytlin, {\it On non-Abelian generalisation of Born-Infeld
action in String Theory}, hep-th 9701125.}
\lref\rBe{E. Bergshoeff, M. Rakowski and E. Sezgin, {\it Higher Derivative
Super Yang-Mills Theories,} Phys. Lett. B185 (1987) 371.}
\lref\rAH{A. Hashimoto and W. Taylor IV, {\it Fluctuation Spectra of Tilted
and Intersecting D Branes from the Born-Infeld Action,} hep-th 9703217.}
\lref\rVaf{B.R. Greene, A. Shapere, C. Vafa and S.T. Yau, {\it Stringy Cosmic
String and noncompact Calabi-Yau manifolds,} Nucl. Phys. B337 (1990) \semi
M. Asano, {\it Stringy Cosmic Strings and Compactifications,} hep-th 9703070.}
\lref\rCV{C. Vafa, {\it Evidence for F Theory,} Nucl. Phys. B469 (1996) 403,
hep-th 9602022.}
\lref\rKeh{A. Kehagias, {\it New type IIB vacua  and their F-theory 
interpretation,} hep-th 9805131.}
\lref\rATT{A.A. Tseytlin, {\it Self Duality of Born-Infeld Action and Dirichlet
Threebrane of Type IIB Superstring Theory,} Nucl. Phys. B469 (1996) 51,
hep-th 9602064.}
\lref\rRadu{C. Ahn, K. Oh and R. Tatar, {\it The Large $N$ Limit of
${\cal N}=1$ Field Theory from F thepry,} hep-th 9808143.}
\lref\rPW{N.D. Lambert and P.C. West, {\it Gauge Fields and M Fivebrane 
Dynamics,} Nucl. Phys. B524 (1998) 141-158, hep-th 9712040.}
\lref\rSD{A. Sen, {\it F Theory and Orientifolds,} Nucl. Phys. B475 (1996)
562, hep-th 9605150\semi
T. Banks, M.R. Douglas and N. Seiberg, {\it Probing F Theory with Multiple
Branes,} Phys. Lett. 387B (1996) 278, hep-th 9605199\semi
O. Aharony, J. Sonnenschein, S. Yankielowicz and S. Theisen, {\it Field Theory 
Questions for String Theory Answers,} Nucl. Phys. 493B (1997) 177, hep-th 
9611222\semi
M.R. Douglas, D.A. Lowe and J.H. Schwarz, {\it Probing F Theory with Multiple
Branes,} Phys. Lett. 394B (1997) 297, hep-th 9612062.}


\Title{BROWN-HET-1155}
{\vbox {\centerline{Higher Derivative Terms from Threebranes}
        \centerline{in F Theory}
}}

\smallskip
\centerline{Robert de Mello Koch and Radu Tatar}
\smallskip
\centerline{\it Department of Physics,}
\centerline{\it Brown University}
\centerline{\it Providence RI, 02912, USA}
\centerline{\tt robert,tatar@het.brown.edu}\bigskip

\medskip

\noindent
The computation of higher derivative corrections to the low energy 
effective actions of ${\cal N}=2$ gauge theories is considered. In
particular, higher derivative corrections are computed for 
four dimensional ${\cal N}=2$
super Yang-Mills theory with gauge group $SU(2)$ and $N_f=4$ 
hypermultiplets in the fundamental representation. The four derivative terms
computed in an approach which realizes the gauge theory as the world volume 
theory of three branes in F therory are in agreement with the field theory 
result. 


\Date{November, 1998}


\newsec{Introduction}

A particularly efficient way to construct the low energy effective action
of a super Yang-Mills theory is to realize the field theory of interest as 
the worldvolume theory of a suitable brane\rHW. 
In this approach, gauge theories
with a reduced number of supersymmetries can be obtained by considering a web
of intersecting branes in type IIA string theory. After lifting to M theory, 
the type IIA web can be realized in terms of a single M theory fivebrane 
wrapping a Riemann surface. The Riemann surface is the Seiberg-Witten curve. 

The limit in which the field theory is realized on the brane world volume
is a low energy limit in which one decouples bulk gravity from the world volume
theory. In addition, the string tension must be taken to be large in order to
decouple open string oscillator excitations. Finally, Kaluza-Klein modes
associated with the brane geometry and the compact eleventh (strong coupling)
dimension have to be decoupled. Quantities in the low energy effective action 
which are constrained by supersymmetry are not sensitive to the limit in which 
they are computed. The more supersymmetry a theory has, the more the low energy
effective action is constrained. 
For the case of ${\cal N}=4$ supersymmetry in four
dimensions, the constraints are so severe that they restrict the form
of four derivative terms in the low energy effective action\rDL. For 
${\cal N}=2$ theories in four dimensions, the constraints due to
supersymmetry imply that the leading low energy effective 
action can be written as an ${\cal N}=2$ superspace chiral integral of a
holomorphic prepotential. For ${\cal N}=1$ supersymmetry in four dimensions, 
supersymmetry constrains the superpotential
to be a holomorphic function of a chiral superfield.
There is an impressive collection of 
holomorphic (BPS) quantities that have been 
computed using the brane approach\rHol. 

The computation of non-holomorphic quantities is more delicate though, and they
are sensitive to the limit in which they are computed. 
Interesting non-holomorphic
quantities include the higher derivtive corrections to the ${\cal N}=2$
super Yang-Mills theory and the K\"ahler potential of the ${\cal N}=1$ theory.  
In \rWest\ these quantities were computed 
using the M theory fivebrane. The results
obtained show a clear quantitative disagreement with what is expected from
the four dimensional gauge theories. This would seem to suggest that although 
the brane approach is a useful tool for computing holomorphic quantities, it
can not be used to compute quantities that are not protected by supersymmetry.
This is unfortunate, since ultimately one would like to get insights into
$QCD$ which is not a supersymmetric theory.

In a recent paper \rM, ${\cal N}=2$ and ${\cal N}=1$ field theories were 
realized as worldvolume theories of Dirichlet 
threebranes moving near sevenbranes, i.e. threebranes in F theory. The authors 
of \rM\ showed that if the number of threebranes 
is large, the geometry can be trusted in the field theory limit, suggesting 
that one could compute non-BPS quantities. The aim of this work is to test
this exciting suggestion in some simple cases.

Specifically, in this article we consider the 
calculation of higher derivative corrections
to the low energy effective action of ${\cal N}=2$ supersymmetric Yang-Mills
field theories in four dimensions. In section two, we begin by reviewing
what is known from field theory about these corrections. In section three,
the computation of these quantities using the Dirichlet fivebrane is
performed for the finite theory with gauge group $SU(2)$ and four massless 
hypermultiplets in the fundamental representation. The computation 
in this case is particularly simple and both the low energy effective action
and the first higher derivative corrections can be computed exactly. The
fivebrane result disagrees with the 
field theory result. 
In section four, we compute the higher derivative 
corrections using threebranes in F theory. The supergravity solution is known,
and the higher derivative corrections can simply be read from an expansion
of the Born-Infeld action. The result is in perfect agreement with
the field theory result. Section five contains a discussion of our results.

\newsec{Field Theory Results}

The low energy effective action of ${\cal N}=2$ super Yang-Mills theory,
when written in ${\cal N}=2$ superspace, has the form

\eqn\First
{S=\int d^{4}xd^{4}\theta {\cal F}(A^{i})+\int 
d^{4}xd^{4}\bar{\theta}\bar{\cal F}(\bar{A}^{i})
+\int d^{4}xd^{4}\theta d^{4}\bar{\theta} {\cal H}(A^{i},\bar{A}^{i}).}

\noindent
The prepotential ${\cal F}$ is a holomorphic function of the abelian 
${\cal N}=2$ chiral vector superfields. This quantity can be computed
directly in field theory using Seiberg-Witten theory\rSW. The real function
${\cal H}(A,\bar{A})$ gives the first non-holomorphic corrections to the
low energy effective action. In general, the exact form of ${\cal H}$ is
not known although several contributions to ${\cal H}$ are known explicitely.
These are the one loop contribution\rOL, 
the two loop contribution\rTL, the one
instanton contribution\rOI\ and the two instanton contribution\rTI. We will be
most interested in the gauge theory with gauge group $SU(2)$ and $N_{f}=4$
massless hypermultiplets in the fundamental representation, which is a 
finite and scale invariant gauge theory. In this case, scale invariance forbids
a normalization scale $\Lambda$ and hence one may be tempted to conclude that
there are no higher loop or
instanton corrections. In this case, because ${\cal H}$ would be one 
loop exact, there are claims that\rOL\

\eqn\Second
{{\cal H}(A,\bar{A})={3\over 256\pi^{2}}ln^{2}\Big(
{A\bar{A}\over \langle A\rangle\langle\bar{A}\rangle}\Big),}

\noindent
in an exact formula. ${\cal H}$ is invariant under the K\"ahler gauge 
transformations

\eqn\Secprime
{{\cal H}(A,\bar{A})\to {\cal H}(A,\bar{A})+f(A)+\bar{f}(\bar{A}),}

\noindent
so that \Second\ is explicitely scale invariant. At this point a comment is in 
order. The absence of a normalization scale $\Lambda$ has been used to argue
that the leading low energy effective action itself does not receive quantum 
corrections. Explicit instanton corrections show that this is not the case.
Thus, the claim that (2.2) is exact is doubtful.

The results we wish to compare with will be expressed in terms of 
components so that we need to find the component expansion of \First. This is
most easily done using an ${\cal N}=1$ superspace notation. The ${\cal N}=1$
chiral superfield contained in $A^{i}$ is denoted by $\Phi^{i}$; the 
${\cal N}=1$ field strength contained in $A^{i}$ is denoted by $W^{i}_{\alpha}$.
The complex scalar appearing in $\Phi^{i}$ is denoted by $\phi^{i}$. Using the
${\cal N}=1$ expansion of \rEx, we find the following four derivative terms for
the scalars $\phi^{i}$\rYOz

\eqn\Third
{\eqalign{S_{4}&=\int d^{4}x\Big[ 2
{\partial^{2}{\cal H}\over \partial\phi_{i}\partial\bar{\phi}_{j}}
(\partial^{m}\partial_{m}\phi^{i})(\partial^{n}\partial_{n}\bar{\phi}^{j})+
{\partial^{3}{\cal H}\over \partial\phi_{i}
\partial\phi_{j}\partial\bar{\phi}_{k}}
(\partial^{m}\phi^{i})(\partial_{m}\phi^{j})
(\partial^{n}\partial_{n}\bar{\phi}^{k})\cr
&+{\partial^{3}{\cal H}\over \partial\bar{\phi}_{i}
\partial\bar{\phi}_{j}\partial\phi_{k}} (\partial^{m}\bar{\phi}^{i})(\partial_{m}\bar{\phi}^{j})
(\partial^{n}\partial_{n}\phi^{k})+
{\partial^{4}{\cal H}\over \partial\phi_{i}\partial\phi_{j}
\partial\bar{\phi}_{k}\partial\bar{\phi}_{l}}
(\partial^{m}\phi^{i})(\partial_{m}\phi^{j})
(\partial^{n}\bar{\phi}^{k})(\partial_{n}\bar{\phi}_{k})\Big].}}

\noindent
Similarily, the kinetic term for the $\phi^{i}$ is

\eqn\ThrdPrm
{S=\int d^{4}x \partial_{m}\phi^{i}\partial^{m}\bar{\phi}^{j}
Im\Big({\partial^{2}{\cal F}\over\partial\phi^{i}\partial\phi^{j}}\Big)
\equiv\int d^{4}x \partial_{m}\phi^{i}\partial^{m}\bar{\phi}^{j}
K_{i\bar{j}}.}

\noindent
In the remaining two sections we will see that the branes provide a form for 
the four-derivative term that is only consistent with the ${\cal N}=2$ field 
theory result after we make certain field redefinitions. The need for these
field redefinitions has been interpreted in \rYOz\ as a consequence of the fact 
that the ${\cal N}=2$ supersymmetry in field theory is realized differently
than it is in the fivebrane field theory. The field equation for $\phi^{i}$ 
reads

\eqn\Fourth
{\partial^{m}\partial_{m}\phi^{i}=-(K_{i\bar{j}})^{-1}
{\partial K_{\bar{j}k}\over \partial\phi_{l}}(\partial^{m}\phi^{k})
(\partial_{m}\phi^{l}).}

\noindent
The field redefinitions that are needed correspond to replacing 
$\partial_{m}\partial^{m}\phi$ in \Third\ with the right hand side of \Fourth.
This leads to the following expression\rYOz\

\eqn\Fifth
{S_{4}=\int d^{4}x \tilde{\cal H}_{ij\bar{k}\bar{l}}(\partial^{m}\phi^{i})
(\partial_{m}\phi^{j})(\partial^{n}\bar{\phi}^{k})
(\partial_{n}\bar{\phi}^{l}),}

\noindent
where

\eqn\Sixth
{\eqalign{\tilde{\cal H}_{ij\bar{k}\bar{l}}&=
{\partial^{4}{\cal H}\over \partial\phi_{i}\partial\phi_{j}
\partial\bar{\phi}_{k}\partial\bar{\phi}_{l}}-
{\partial^{3}{\cal H}\over \partial\phi_{i}\partial\phi_{j}
\partial\bar{\phi}_{p}}
(K_{\bar{p}q})^{-1}{\partial K_{q\bar{k}}\over\partial\bar{\phi}^{l}}-
{\partial K_{j\bar{p}}\over \partial\phi^{i}}(K_{\bar{p}q})^{-1}
{\partial^{3}{\cal H}\over \partial\phi_{q}
\partial\bar{\phi}_{k}\partial\bar{\phi}_{l}}\cr
&+2{\partial K_{j\bar{p}}\over\partial\phi^{i}}(K_{\bar{p}q})^{-1}
{\partial^{2}{\cal H}\over \partial\phi_{q}\partial\bar{\phi}_{r}}
(K_{\bar{r}s})^{-1}{\partial K_{s\bar{k}}\over\partial\bar{\phi}^{l}}.}}

\noindent
Using the explicit expressions (valid for $N_c=2$ and $N_f=4$ massless
hypermultiplets in the fundamental representation)

\eqn\SevPrme
{K_{u\bar{u}}={Im(\tau)\over 8\sqrt{u\bar{u}}},\quad u={1\over 2}A^{2},
\quad \tau={\theta\over\pi}+{8\pi i\over g^{2}},}

\noindent
and the formula \Second\ for ${\cal H},$ we finally find

\eqn\Seventh
{S=\int d^{4}x(\partial^{m}u\partial_{m}u)
(\partial^{n}\bar{u}\partial_{n}\bar{u}){3\over 2^{8}\pi^{2}u^{2}\bar{u}^{2}}.}

\noindent
The formulas (2.9) and (2.10) do not include instanton corrections.
Before leaving this section, we note that in the pure gauge case, the one loop
results for $SU(2)$ are

\eqn\Eigth
{K_{u\bar{u}}\sim {log(16u\bar{u}/\Lambda^{4})\over\sqrt{u\bar{u}}},\quad
u={1\over 2}A^{2},\quad H(A,\bar{A})\sim log\Big({A\over\Lambda}\Big)
log\Big({\bar{A}\over\Lambda}\Big).}

\noindent
Thus, the semiclassical four derivative term reads\rYOz\ 

\eqn\KJHKJ
{S=\int d^{4}x (\partial^{m}u\partial_{m}u)
(\partial^{n}\bar{u}\partial_{n}\bar{u})
{8+4log\Big({16u\bar{u}\over\Lambda^{4}}\Big)
+\Big[log\Big({16u\bar{u}\over\Lambda^{4}}\Big)\Big]^{2}\over
u^{2}\bar{u}^{2}\Big[log\Big({16u\bar{u}\over\Lambda^{4}}\Big)\Big]^{2}}.}

\noindent
Thus, in the large $u$ (semiclassical) region, the fall off
of the four derivative correction is again $|u|^{-4}$.

\newsec{The Fivebrane Description}

In this section we describe the fivebrane description of the ${\cal N}=2$
super Yang-Mills theory with gauge group $SU(2)$ and $N_{f}=4$ 
massless hypermultiplets
in the fundamental representation. The relevant brane configuration is
realized in type IIA string theory. It consists of two parallel Neveu-Schwarz
fivebranes, with world volume coordinates 
$x^{0},x^{1},x^{2},x^{3},x^{4},x^{5}$. These two fivebranes are separated by a finite distance in the $x^{6}$ direction
and two Dirichlet fourbranes with world
volume coordinates $x^{0},x^{1},x^{2},x^{3},x^{6}$ are suspended between the
two fivebranes. There are four semi infinite fourbranes with world volume
coordinates  $x^{0},x^{1},x^{2},x^{3},x^{6}$. Two semi infinite fourbranes
stretch from $x^{6}=-\infty$ to the left most fivebrane and another two semi
infinite fourbranes stretch from $x^{6}=\infty$ to the right most fivebrane.
We will take the $x^{1}$ and $x^{7}$ directions to be finite.
Using the arguments given in \rJR, this type IIA brane configuration can be
mapped into a single Dirichlet fivebrane in IIB string theory as follows:
Lifting this type IIA configuration to M theory, we obtain a single M theory
fivebrane wrapped on the Seiberg-Witten curve\rEW. If we now return to IIA 
string theory, interpreting $x^{1}$ as the direction which grows at strong
coupling, we obtain a single Dirichlet 
fourbrane wrapping the Seiberg-Witten curve.
Finally, performing a T duality along the $x^{7}$ direction, we obtain a
single Dirichlet fivebrane in type IIB string theory. The Seiberg-Witten curve
for the above brane configuration takes the form\rEW\

\eqn\first
{\eqalign{v^{2}t^{2}-2B(v)t+ev^{2}=0,\quad B(v)=v^{2}+u,\cr
t=exp(-s/R_{7})=exp(-(x^{6}+ix^{7})/R_{7}),\quad v=x^{4}+ix^{5}.}}

\noindent
The Dirichlet fivebrane has world volume coordinates
$x^{0},x^{1},x^{2},x^{3},x^{6},x^{7}$. The low energy world volume desciption
of this Dirichlet fivebrane is given by the following $5+1$ dimensional 
Yang-Mills theory

\eqn\BosonicPart
{{\cal L}=
Tr\Big(F_{\mu\nu}F^{\mu\nu}+D_{\mu}X^{I}D^{\mu}X^{I}+\big[
X^{I},X^{J}\big]^{2}\Big),}

\noindent
where $I=4,5,8,9,$ $\mu,\nu =0,1,2,3,6,7$ and only the bosonic part of the 
Lagrangian is shown. The $X^{I}$ are $2\times 2$ dimensional matrices. The 
classical fivebrane solution\rJR\
is given by taking $X^{4}$ and $X^{5}$ diagonal
and setting all other fields to zero. It is convenient to assemble the
eigenvalues $x_{i}^{4}$ and $x_{i}^{5}$ of $X^{4}$ and $X^{5}$ into the single
complex number $v_{i}=x^{4}_{i}+ix_{i}^{5}$. The complex numbers $v_{i}$ are
now identified with the roots of the Seiberg-Witten curve \first. 
In this way, the Higgs fields trace out the curve described in \first\ as
the worldvolume coordinates vary so that we do indeed obtain a fivebrane
wrapped on the Seiberg-Witen curve. For the case 
that we study here, the roots $v_{i}$ are given by

\eqn\second
{v_{1,2}=\pm\sqrt{2u}\sqrt{t\over t^{2}-2t+e}.}

\noindent
Notice that the sum of roots vanishes so that the Higgs fields can be expanded
in the Lie algebra of $SU(2)$ as expected. The terms in the action 
\BosonicPart\ which give rise to the scalar kinetic term of the four
dimensional field theory are ($m=0,1,2,3$, $Y=X^{4}+iX^{5}$)

\eqn\RelTerms
{{\cal L}_{kin}=\int d^{2}s Tr\Big(\partial_{m}Y\partial^{m}
Y^{\dagger}\Big)=\int d^{2}s \partial_{m}v_{i}\partial^{m}\bar{v}_{i},}

\noindent
The $u$ dependence of the action can be extracted without performing any
explicit integrals

\eqn\third
{S=\int d^{4}x{\partial_{m}u\partial^{m}\bar{u}\over 8\sqrt{u\bar{u}}}
Im(\tau),\quad Im(\tau )=4\int d^{2}s 
\sqrt{t\bar{t}\over(t^{2}-2t+e)(\bar{t}^{2}-2\bar{t}+e)}.}

\noindent
A few comments are in order. The above $u$ dependence of the effective action 
shows that $a\sim\sqrt{u}$. This is the expected result. It would be wrong to
conclude that the effective action action has not received any perturbative or
instanton corrections. In the case of finite gauge theories, there are both
loop and instanton corrections\rDor. These corrections enter in the relation 
between the parameters in the fivebrane curve and parameters in the field
theory. Note however, that independently of this relation, $\tau$ is a 
constant. The ease with which we evaluated the $u$ dependence
of the low energy effective action is a direct consequence of this.

The higher derivative corrections to the super Yang-Mills theory are expected
to arise from a non Abelian Born-Infeld action. An explicit form for this
action has been suggested by Tseytlin \rTS. Although there have been some 
questions regarding the validity of this action \rAH, 
our solutions are diagonal 
matrices and we do not expect further corrections, which presumably probe the
non-Abelian structure of the solution, to affect our result. For that reason, 
we will consider the action

\eqn\NABi
{\eqalign{S_{p}=T_{p}\int d^{p+1}xSTr&\Big[
\sqrt{-det(\eta_{rs}+D_{r}X_{a}(\delta_{ab}-iT\big[X_{a},X_{b}\big])^{-1}
D_{s}X_{b}+T^{-1}F_{mn})}\cr
\times&\sqrt{det(\delta_{ab}-iT\big[X_{a},X_{b}\big])}\Big],}}

\noindent
where $T_{p}$ is the $p$-brane tension, $T^{-1}=2\pi\alpha'$ and the
symmetrized $STr$ is defined by

\eqn\STrcd
{STr(A_{1}...A_{n})={1\over n!}Tr\Big(A_{1}...A_{n}+
{\tt all}\quad{\tt permutations}\Big).}

\noindent
At low energy, we regain the super Yang-Mills description from this action.
The above action can be expanded as follows

\eqn\ExpndNBI
{\eqalign{S&=Tr(L)+{1\over 2}Tr(M_{r}^{r}L)+{1\over 8}Tr(M_{r}^{r}M_{s}^{s}L)
-{1\over 4}Tr(M_{rs}M^{sr}L)\cr
L&=\sqrt{det(\delta_{ab}-iT\big[X_{a},X_{b}\big]}\cr
M_{rs}&=D_{r}X_{a}(\delta_{ab}-iT\big[X_{a},X_{b}\big])^{-1}
D_{s}X_{b}+T^{-1}F_{rs}).}}

\noindent
Although we have only considered the bosonic piece of the fivebrane action,
it is interesting to note that a supersymmetric extension of \ExpndNBI\ has
been constructed in \rBe. Inserting the classical solution, 
the higher derivative
corrections take the form

\eqn\HDCrns
{S\sim\int d^{4}xTr(\partial_{m}Y\partial^{m}Y\partial_{n}\bar{Y}
\partial^{n}\bar{Y})\sim
\int d^{4}x\partial_{m}u\partial^{m}u\partial_{n}\bar{u}
\partial^{n}\bar{u}{1\over u\bar{u}}.}

\noindent
Notice that the higher derivative corrections obtained from the fivebrane have
the same structure as the higher derivative corrections computed in field 
theory. It is clear however that the $u$ dependence of the four derivative 
terms disagree with the field theory result. The $u$ dependence of the above
result is in perfect agreement with the $u$ dependence obtained in \rPW,
where the higher derivative corrections from a fivebrane wrapping the
Seiberg-Witten curve corresponding to pure $SU(2)$ ${\cal N}=2$ gauge
theory were estimated.

This discrepancy between the field theory result and the fivebrane result is
not unexpected, as we now explain. The ${\cal N}=2$ super Yang-Mills theory
is expected to arise from the IIA brane configuration at low energy and weak
string coupling. The analysis we have performed for the Dirichlet fivebrane is 
valid at weak string coupling and low energy. Thus, for the analysis of this 
section to be appliable to the field theory, we need to verify that the weak
coupling low energy description of the Dirichlet fivebrane is dual to the 
weak coupling low energy description of the IIA configuration. The 
results of \rJR\ show that the low energy weakly coupled type IIA description 
is dual to a strong coupling low energy description of the type IIB Dirichlet
fivebrane. Thus, there is no reason to expect that the higher derivative 
corrections computed using the fivebrane should be related to the higher 
derivative corrections of the ${\cal N}=2$ super Yang-Mills theory. This is a
good example showing that non-holomorphic corrections are sensitive to the 
limit in which they are computed.

\newsec{Threebrane in F Theory}

We begin by reviewing the supergravity solution for 
threebranes moving in a sevenbrane
background given in \rM\foot{This solution has also appeared in \rVaf. For 
additional work on the large $N$ limit of field theory from threebranes in F
theory see\rRadu.}. We start from a solution for the 
sevenbranes by themselves.
The NSNS two form and RR two and four forms are set to zero. 
This leaves the metric
and the dilaton from the NSNS 
sector and the axion from the RR sector. It is
convenient to combine the dilaton and axion into a single complex coupling
$\tau=\tau_{1}+i\tau_{2}=\chi+ie^{-\phi}$. The parameter $\tau$ is the modular
parameter of the elliptic fiber of the 
F theory\rCV\ compactification. Introduce the 
complex coordinate $z=x^{8}+ix^{9}$. In terms of $z$ we take the following 
ansatz for the metric

\eqn\frst
{ds^{2}=e^{\phi(z,\bar{z})}dzd\bar{z}+dx_{7}^{2}+...+dx^{2}_{1}-dx_{0}^{2}.}

\noindent
This ansatz is for a sevenbrane with worldvolume coordinates
$x^{0},x^{1},x^{2},x^{3},x^{4},x^{5},x^{6},x^{7}$.
With this ansatz, the type IIB supergravity equations reduce to\rVaf\

\eqn\scnd
{\eqalign{
\partial\bar{\partial}\tau &=
{2\partial\tau\bar{\partial}\bar{\tau}\over\bar{\tau}-\tau}\cr
\partial\bar{\partial}\phi &=
{\partial\tau\bar{\partial}\bar{\tau}\over(\bar{\tau}-\tau)^{2}}.}}

\noindent
The sevenbrane background of relevence for the ${\cal N}=2$ field theory is
obtained by identifying $\tau$ with the effective gauge coupling constant.
This implies that $\tau=\tau(z)$ so that the first equation in \scnd\ is
automatically satisfied. The general solution to the second equation in
\scnd\ is

\eqn\thrd
{\phi (z,\bar{z})=log\tau_{2} +F(z)+\bar{F}(\bar{z}).}

\noindent
The functions $F(z)$ and $\bar{F}(\bar{z})$ should be chosen in order that
\frst\ yields a sensible metric. For the case that we are considering (i.e.
constant $\tau$), the explicit form for the metric transverse to
the sevenbranes is\rVaf\

\eqn\htgdgf
{ds^{2}=e^{\phi (z,\bar{z})}dzd\bar{z}=\tau_{2}|da|^{2},}

\noindent
where $a$ is the quantity that appears in the Seiberg-Witten solution\rM.
This specifies the solution for the sevenbranes by themselves.

Next, following \rM\ we introduce threebranes into the problem\foot{See also
\rKeh\ where this solution was independently discovered.}. The world volume
coordinates of the threebranes are $x^{0},x^{1},x^{2},x^{3}$. One obtains a
valid solution by making the following ansatz for the metric

\eqn\frth
{ds^2 = f^{-1/2}dx_{\parallel}^2 + f^{1/2}{g}_{ij}dx^{i}dx^{j}}

\noindent
and the following ansatz for the self-dual 5-form field strength

\eqn\ffth
{F_{0123i}  =  -{{1}\over{4}}\partial_{i}f^{-1}~. }

\noindent
This solution corresponds to introducing $N$ coincident threebranes. 
The complex field $\tau$ is unchanged. Inserting the above ansatz into the
IIB supergravity equations of motion, one finds that $f$ satisfies the
following equation of motion\rM\

\eqn\sxh
{ {1 \over \sqrt{ g} } 
\partial_{i}(\sqrt{ g}{ g}^{ij}\partial_j f)=
- (2 \pi)^4 N { \delta^6(x-x^0) \over \sqrt{\ g}. }}

\noindent
In the limit that $N\to\infty$ the curvature becomes small almost everywhere
and the supergravity solution can be used to reliably compute quantities in
the field theory limit as explained in\rM. A sensitive test of this claim
is the computation of higher derivative corrections performed below. 

To obtain information about the ${\cal N}=2$ super Yang-Mills theory, we now
consider a threebrane separated from the rest of 
the threebranes. The dynamics of this threebrane 
probe is given by a Born-Infeld action in the above supergravity
background. The leading low energy effective action plus four derivative terms 
for the scalars are thus obtained by expanding the action\rATT\

\eqn\svnth
{\eqalign{S&={T_{3}\over 2}
\int d^{4}x\Big[\sqrt{det(G_{mn}+e^{-{1\over 2}\phi}F_{mn})}+\chi
F\wedge F\Big]\cr
&={T_{3}\over 2}\int\Big(\tau_{2}F^{2}+\tau_{1}F\wedge F+
e^{\phi(z,\bar{z})}\partial_{m} z\partial^{m} \bar{z}
+fe^{2\phi(z,\bar{z})}
\partial_{m} z\partial^{m} z\partial_{n}\bar{z}\partial^{n}\bar{z}\Big).}}

\noindent
where $z=x^{8}+ix^{9}$ and $x^{i}$ with $i=4,5,6,7$ have been set to zero.
It is clear that the low energy effective action of the threebrane probe is 
the same as the exact solution of the corresponding low-energy field 
theories\rSD.
Note once again that the brane answer for the four derivative terms has the 
same general structure as the four derivative terms computed in field theory.

We are now ready to return to the super Yang-Mills theory with $N_c=2$ and
$N_f=4$. In this case, the supergravity solution can be determined exactly\rM.
The metric transverse to the sevenbranes takes the form given in (2.2).
In terms of $a$ the solution of \sxh\ reads

\eqn\eght
{\tau={\theta\over\pi}+
{8\pi i\over g^{2}},\quad f={Nc_{1}\over (\tau_{2}|a|^{2}+y^{2})^{2}}.}

\noindent
with $c_{1}$ a constant which can be fixed using \sxh.
The coordinate $y$ is transverse to the threebranes but parallel to the 
sevenbranes. The coordinate $a$ is transverse to both the seven branes and the 
three branes. The $N$ threebranes are at $y=a=0$. The probe threebrane is
at $y=0$ and at some $a\ne 0$. Moving the probe in the $a$ direction
corresponds to moving in the moduli space of the ${\cal N}=2$ field theory.  
Evaluating the probe action \svnth\ at this solution, we find

\eqn\nnth
{S={T_{3}\over 2}
\int d^{4}x\Big(\tau_{2}\partial_{n}a\partial^{n}\bar{a}
+Nc_{1}{1\over a^{2}\bar{a}^{2}}
\partial_{m} a\partial^{m} a\partial_{n}\bar{a}\partial^{n}\bar{a}
\Big)}

\noindent
for the scalar fields in the probe action.
This is in perfect {\it quantitative} agreement with the field theory
results. 
Note that the present computation does seem to test the
coefficient in front of the four derivative term, as we now explain.
The relation between the Higgs expectation value of the field theory and 
the corresponding threebrane coordinate allows the introduction
of one multiplicative constant $a=ca_{SW}$ for any constant $c$. Since the two
terms in the low energy effective action scale with different powers of
$c$, their relative normalization can be fixed to the field theory prediction
by a judicious choice of $c$. The tension of the threebrane introduces an
overall constant which can then be fixed so that the overall normalisation
of the probe action and the field theory action agree. 
 
\newsec{Discussion}

In this letter we have considered the computation of non-holomorphic
quantities using the Dirichlet fivebrane and threebranes in F theory. The
results obtained using the fivebrane disagree with the field theory results.
This disagreement could be traced back to the fact that the description of
the Dirichlet fivebrane was not valid in the limit in which field theory
is expected to emerge. This clearly illustrates the fact that the four
derivative terms are not constrained by supersymmetry.
The results obtained using threebranes in F theory are 
in perfect agreement with field theory. The good agreement in this case is
due to the fact 
that the supergravity solution is valid in the field theory limit,
if one takes a large number of three branes. To the best of our
knowledge, this is the first time that corrections to a field theory
which are not protected by supersymetry, have been computed using a brane
approach. This suggests that the method
derived in \rM\ provides a reliable approach to the computation of 
non-holomorphic corrections. This is an important result
because these quantities can not, at present, be computed directly in the
field theory.

{\it Acknowledgements} RdMK is supported by a South African FRD post doctoral
bursary. We would like to thank J.P. Rodrigues for numerous discussions on
higher derivative corrections from fivebranes and for helpful comments on
version 1.

\listrefs
\vfill\eject
\bye